# The SARS-CoV-2 Spike Protein is vulnerable to moderate electric fields

Claudia R. Arbeitman [1,2,3,*], Pablo Rojas[1,*], Pedro Ojeda-May[4] & Martin E. Garcia[1,**]

1 Theoretical Physics and Center of Interdisciplinary Nanostructure Science and Technology, FB10, Universität Kassel, Heinrich-Plett-Str. 40, 34132 Kassel, Germany.

2 CONICET Consejo Nacional de Investigaciones Científicas y Técnicas. Godoy Cruz 2290 C1425FQB, Buenos Aires, Argentina.

3 GIBIO-Universidad Tecnológica Nacional-Facultad Regional Buenos Aires. Medrano 951 C1179AAQ, Buenos Aires, Argentina.

4 High Performance Computing Centre North (HPC2N), Umeå University, Umeå SE-90187, Sweden.

* These authors contributed equally

** corresponding author

**Most of the ongoing projects aimed at the development of specific therapies and vaccines against COVID-19 use the SARS-CoV-2 spike (S) protein as the main target [1-3]. The binding of the spike protein with the ACE2 receptor (ACE2) of the host cell constitutes the first and key step for virus entry. During this process, the receptor binding domain (RBD) of the S protein plays an essential role, since it contains the receptor binding motif (RBM), responsible for the docking to the receptor. So far, mostly biochemical methods are being tested in order to prevent binding of the virus to ACE2 [4]. Here we show, with the help of atomistic simulations, that external electric fields of easily achievable and moderate strengths can dramatically destabilise the S protein, inducing long-lasting structural damage. One striking field-induced conformational change occurs at the level of the recognition loop L3 of the RBD where two parallel beta sheets, believed to be responsible for a high affinity to ACE2 [5], undergo a change into an unstructured coil, which exhibits almost no binding possibilities to the ACE2 receptor (Figure 1a). Remarkably, while the structural flexibility of S allows the virus to improve its probability of entering the cell, it is also the origin of the surprising vulnerability of S upon application of electric fields of strengths at least two orders of magnitude smaller than those required for damaging most proteins. Our findings suggest the existence of a clean physical method to weaken the SARS-CoV-2 virus without further biochemical processing. Moreover, the effect could be used for infection prevention purposes and also to develop technologies for in-vitro structural manipulation of S. Since the method is largely**

**unspecific, it can be suitable for application to mutations in S, to other proteins of SARS-CoV-2 and in general to membrane proteins of other virus types.**

SARS-CoV-2, the agent responsible for the outbreak of COVID-19, is an enveloped virus that utilises its surface glycoprotein spike (S) to bind to the host cell membrane through an angiotensin-converting enzyme (ACE2) receptor [6,7]. Most of the efforts to develop therapies and vaccines against COVID-19 [8–11] aim at either decreasing the stability or exploiting some of the structural features of the S protein, since it triggers immune responses and plays an essential role in the virus ability to infect the host [12,13]. The S protein is a homotrimer whose protomers are composed of two functional subunits, S1 and S2, which are responsible for the correct receptor binding/fusion of the viral and cellular membranes, respectively. The S1 subunit contains the receptor-binding domain (RBD), which has been found to switch stochastically between a closed ("down") state, in which the receptor binding motif (RBM) is hidden, and an open ("up") state that exposes the RBM thus enabling the interaction and binding with the peptidase domain of ACE2 [6,14]. Greater preponderance of the up conformations on mutated S proteins have been linked to higher infectivity, but at the expense of more vulnerability to neutralising antibodies [15]. Distal mutations from the binding region have also been found to affect structural stability of the S protein and its affinity to ACE2, which indicates that a correct spatial arrangement of the RBM residues participating in the binding to the receptor is crucial [16]. In addition to this, abundant N-linked glycans decorating the S protein have been found to be involved in both the stability/folding of the protein in its prefusion conformation and also in controlling the access of host proteases and antibodies, which provides the virus with support in bypassing the host's immune response [17]. Activation of the membrane fusion is achieved by cleaving the S protein by host proteases at the specific sites located at the boundary between the S1 and S2 subunits. This cleavage triggers a rearrangement from the metastable prefusion state into the post-fusion conformation, and the interruption of this process has been shown to prevent viral fusion [18-22]. Altogether, the structure and dynamics of the S protein have been suggested to be the result of a finely tuned balance of affinity to ACE2, stability, and exposure of the RBM [6,7,15,23,24]. Thus, finding new ways to disrupt this balance might result in an additional set of tools to control the virus and therefore the pandemic.

It has been theoretically predicted and experimentally demonstrated that static and time-dependent electric fields (EFs) are capable of inducing conformational changes [25-30] or even irreversible damage in proteins [31-33]. The fact that extremely intense EFs of strengths larger than 1Volt per nanometre ($10^9$ V m$^{-1}$) can denature entire proteins and even break chemical bonds is trivial and of little biological relevance. However, the effect of moderate fields is subtler and can be understood in terms of the interactions of the EFs with the permanent dipoles located in the backbone structure and with the additional flexible dipoles on the protein side chains (see Figure 1d). For instance, under the action of an EF, the electric dipole moments can be reoriented along the field direction in order to minimise the electrostatic energy. On the other hand, a rearrangement of the dipoles can cost conformational energy due to the loss of hydrogen bonds. As a result of the balance between conformational and electrostatic energies along with entropic contributions, the protein can undergo a significant conformational change. In this work, we show, via molecular dynamics (MD) simulations, that EFs of moderate intensities cause damage on the S protein that affects its interaction with ACE2, potentially making SARS-Cov-2 less infectious. These results pave the way to a range of possible applications of EFs to control structural changes in virions with SARS-CoV-2 being one of the multiple targets.

## Moderate electric fields induce global long-lasting structural changes in the spike glycoprotein of SARS-CoV-2

We studied the effect of external EFs on the secondary and tertiary structures of the S protein by performing molecular dynamics simulations. We first considered a representative selected segment of S from a protomer in conformation "up" between residues 319 and 686. This segment corresponds to a part of the S1 subunit and includes the whole RBD, the subdomains SD1 and SD2, and the interface between S1 and S2 (Figure 1b). Previous computational [34-36] and experimental [37,38] works on the isolated S protein and also on the related S protein in SARS-CoV-1 have shown that a standalone segment comprising the RBD and its neighbouring subsequence preserves the local structure and therefore the dynamical and biochemical properties that it shows in the entire protein complex. Based on this knowledge we simulated a restricted spatial domain without loss of generality (see also Methods). To construct the initial protein conformation for the simulations, we used the cryo-EM structure PDB ID 6VSB obtained from the Protein Data Bank [6] and completed the missing residues (see Methods). The first production run was aimed at thermalising the system in the absence of EF (no-EF run) in order to bring the protein to thermodynamic equilibrium at 30 degree Celsius. An estimate of the free energy profile (see Methods) reveals that motion during the thermalisation run was confined to a single free energy basin (Figure 2b). This result indicates that the initial conformation fetched from the experimentally obtained segment was close to a stable equilibrium folding state, and the thermalisation run merely helped to relax the remaining structural stress.

Next, using the thermalised structure as initial state, we carried out simulations on the S protein fragment under the action of an EF during 700 ns. We performed different runs (EF-on runs) corresponding to different EF intensities. The field intensities were selected to span a range between $10^4$ V m$^{-1}$ and $10^7$ V m$^{-1}$ in order to cover both low and moderate intensities that are not incompatible with living organisms and can even exist inside cells [39,40]. Only for the sake of comparison, we also performed a short simulation for an unrealistically high intensity ($10^9$ V m$^{-1}$). In all cases, trajectories display an elongation of the protein as a result of the alignment of permanent local dipoles and displacement of charges parallel to the EF (Figure 2). For the extreme case of EF=$10^9$ V m$^{-1}$, the structural changes are so dramatic that a complete loss of the secondary and tertiary structures of the protein occurs within few ns (see Extended Data, Fig.1). In contrast, for low to moderate intensities (EF < $10^7$ V m$^{-1}$), the field-induced structural changes in the S protein are characterised by a transition to a *new stable* conformation within a few hundreds of nanoseconds (Figure 2a, EF-off). For EF= $10^7$ V m$^{-1}$, the protein structure undergoes, besides the above-mentioned transition, an additional structural change in the region between the SD1 and SD2 subdomains, as shown in Figures 2a and 2d and discussed below. The conformational changes of S upon EF application are reflected in the time evolution of the root mean square displacements (RMSD) of the protein backbone relative to the starting structures (Figure 2a). A transition from a conformation exhibiting stable RMSD-values below ~0.5 nm to a new stable structure showing small RMSD-oscillations around a larger value occurs within the first 200 ns. For EF=$10^7$ V m$^{-1}$ the unfolding between SD1 and SD2 completes only after 500 ns showing a shift to higher values of RMSD around ~1.9 nm. Taken together, these results suggest that EFs modify the free energy balance enabling the protein to overcome barriers which in turn results in a shifted conformational ensemble.

To further assess the stability of the new conformations adopted by S under the external fields, we switched off the EF and continued the simulation in absence of fields (EF-off run). Typically, not more than 200 ns were needed for each EF-off run, since for all EF intensities the protein displayed a restricted motion around the structure left after the field application, as revealed by the RMSD plots. For instance, the unfolding of the region between SD1 and SD2 observed under EF=$10^7$ V m$^{-1}$ remains unaltered after switching off the EF. For all studied EF intensities, estimated free energy plots (see Figure 2b and Extended Data Fig.2) confirmed the existence of a new stable minimum and interestingly, they also show that an energy barrier prevents a transition back to the original conformation (Extended Data Fig. 2). To visualise the relevant conformational changes, we conducted a principal component analysis (PCA) on the trajectories of the EF-on runs. We considered a subspace spanned by the two most relevant principal components for the run at EF=$10^7$ V m$^{-1}$ (see Methods). Then, we projected the trajectories for all EF-off runs onto that plane (Figure 2d). Under the action of EFs of different intensities the protein goes through different paths in the phase space. The final conformation after each EF-on run depends on the field intensity. After EF switch-off and during the EF-off runs the protein structure remains around the EF induced new conformations. No return to the initial structures was observed. This can be clearly seen in Figure 2d, where points corresponding to the trajectories cluster around the final states with almost non-overlapping regions in the reduced phase space (Figure 2d). The qualitative difference of the states explored by the protein during transient and relaxation driven by static EFs only differing in magnitude suggests the possibility of designing EFs to achieve a predetermined structural change. Altogether, these results provide further evidence that the EF induced conformational changes in the S protein are long lasting and do not reverse upon removal of EF.

Considering clustered residues as rigid bodies, global conformational changes can be roughly described by the angles formed by the vectors connecting centroids of the domains of interest [41]. We quantified the influence of EF on the unfolding process observed in the region between SD1 and SD2 (see Figure 1b) by the angle $\theta$ formed between the vectors that connect the centroids of RBD, SD1 and SD2 (Figure 2e). The difference $\Delta\theta$ between the average values before and after EF application increases monotonously with the EF intensity. The shapes of the distributions during the EF-off runs (Figure 2c) are clearly clustered with partial overlaps. It is, however, important to point out at this stage, that the angle-shifts for increasing EF do not merely mean quantitatively different versions of the same structural change. Instead, the average value of and its distributions (see Figure 2e) are determined by different EF induced tertiary structure rearrangements on other parts of the protein not captured by the coordinate $\Delta\theta$. This means that, for each EF, the accessible angles are constrained by barriers of different origin. This confirms our previous observation that conformational changes for different intensities are qualitatively different in the PCA space and not just amplified versions of the motions at lower intensities (Figure 2c). Note, that this effect is similar to that induced by mutations. It was, for instance, reported that certain mutations are able to change the statistics of accessed states at distal sites, namely favouring "up" conformations of RBD by mutation of residue 614 [38,43]. These results show the potential of EF of different strengths to induce changes in S that do not only affect the overall configuration, but also reshape local interactions.

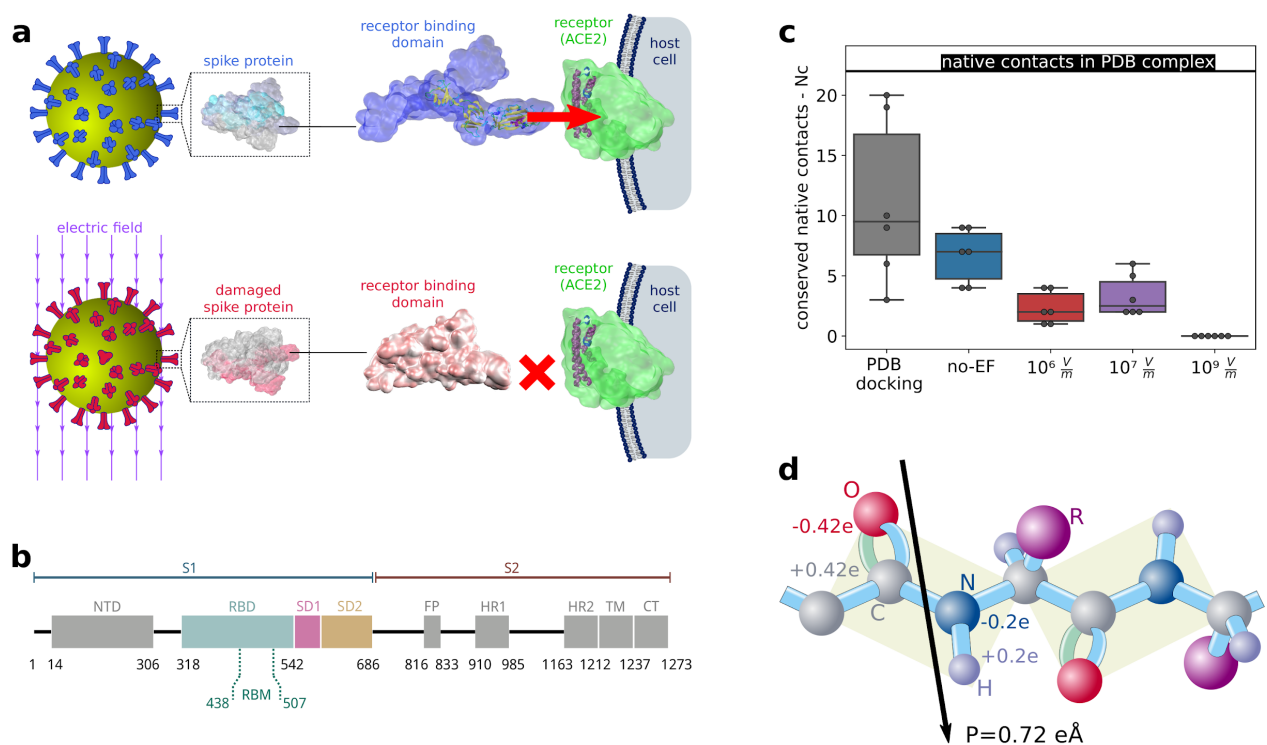

**Figure 1**: External electric fields affect the attachment of SARS-Cov-2 to the host cell. **a,** Virus entry into the cell is mediated by the recognition between the spike glycoprotein (S protein) present in the virus envelope and the angiotensin-converting-enzyme receptor (ACE2) of the host cell membrane. The binding between the S protein and ACE2 can be altered when external electric fields induce drastic conformational changes and damage in the S protein. **b,** Sequence of the S protein (PDB IDs: 6VSB and 6M0J [6, 20]). Highlighted in red square is the segment used in this study. **c,** Conserved number of native contacts ($N_C$) between residues of S and ACE2 for different magnitudes of the EF strength. $N_C$ is maximal for native S protein. Very strong electric fields ($10^9$ V/m) disable the protein by largely deforming its shape, leaving a structure which is unrecognised by ACE2 ($N_C = 0$). Moderate electric fields, which can be induced by available industrial or laboratory devices [63], strongly reduce $N_C$ and are therefore candidates to decrease the affinity of S to ACE2

and, consequently, the infectivity of the virus. **d,** Changes in the structural conformation of proteins under EF are driven by reorientation of electric dipoles.

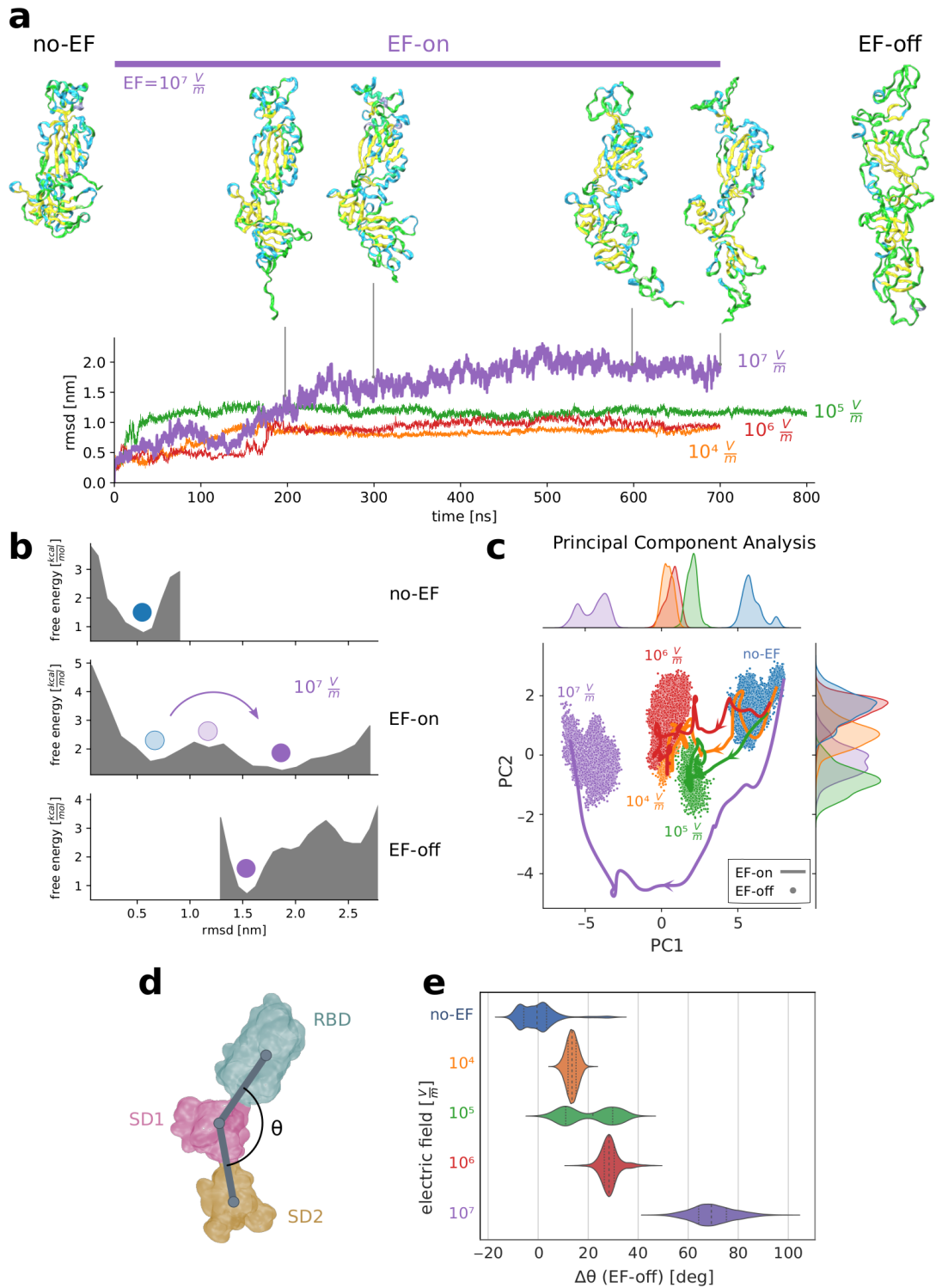

**Figure 2**: Electric fields are able to induce global conformational changes in the spike glycoprotein, affecting the stability of folding states. **a-b,** EF driven major shape changes occur in the different subunits and between subunits of the S protein. a) Snapshots of the studied fragment of S under an EF of $10^7$ V m$^{-1}$ at 0ns (initial thermalised stable conformation), 200ns, 300ns, 600ns and 700ns, and after EF-off (see text) dynamics during 200ns. The orientation of the protein is the same in all figures. Trajectories for different electric field strengths are quantified through the root-mean-square-displacement (RMSD) with respect to the initial structure. Snapshots in **a** correspond to different times along the EF-on trajectory. **b,** Electric fields modify the free energy landscape enabling the protein to overcome potential barriers. Estimated free energy landscape along the thermalisation (no-EF-), EF-on- and EF-off trajectories (see Methods). The blue and the light blue dots identify the energy minimum of the initial structure before and during EF application, respectively. Purple dots correspond to the new minimum reached under the EF, which remains stable after switching off the EF. **c,** Principal component analysis (PCA) reveals the existence and stable nature of new states after EF application (see Methods). Discretised trajectories of the EF-on and EF-off runs projected onto a plane defined by the two principal components (PC1 and PC2). Curves on the upper and right axis show the density of points along PC1 and PC2, respectively. Once the S protein has found a new equilibrium basin, which is different for each EF intensity, no return to the initial state occurs after switch-off of the EF. **d,** Field-induced conformational states can be characterised by the angles formed by the vectors connecting the centroids of clustered residues. **e,** Violin plot of the distributions of the shift $\Delta\theta$ of the angle $\theta$ for different field intensities (EF-off runs) with respect to a no-EF representative structure. $\Delta\theta$ is suitable to describe the unfolding of the domain SD2 observed in **a**. EF intensities are color-coded equally for all sub-figures.

## Electric fields strongly affect the binding of the spike protein with the ACE2 receptor

To study how EF specifically disturb the stability of the RBD and, in particular, of residues that are vital to the local interaction with ACE2, we performed MD simulations using a model for the structure of the unbound RBD obtained from a crystallographic experimental structure of the RBD-ACE2 complex (PDB ID 6M0J) [20]. We first conducted thermalisation simulations under EF = 0 V m$^{-1}$ (no-EF), which showed that the tertiary structure is preserved compared to the initial crystal structure (see Figure 3). Previous studies [43,44] have shown that the RBD can be described as being formed by a core and the RBM. The loop 3 (L3), between residues Tyr470 and Pro491, is one of the four loops comprising the RBM, and has been demonstrated to play a key role in the interaction of S with ACE2 [20]. The presence of two small $\beta$-strands in the fragments Cys488–Tyr489 and Tyr473–Gln474 of L3 was shown to be one of the reasons for the enhanced affinity of SARS-CoV-2 with ACE2, which is 15 to 20 times larger than the affinity of SARS-CoV-1, whose S protein exhibits a L3-loop without $\beta$-sheets[6,41]. The larger affinity to ACE2 makes SARS-CoV-2 much more infectious than SARS-CoV-1[5,45]. The above-mentioned beta-strands in L3 remained intact along the thermalisation simulations, in agreement with previous studies showing that the greater rigidity of the $\beta$-sheets increases the stability of L3 in SARS-CoV-2 as compared to the unstructured L3 in SARS-CoV-1 [5,45-49]. Furthermore, the rest of the secondary structure of the RBD was observed to be stable during the thermalisation run (Figure 3a, left structure). These results are consistent with current studies providing evidence for the stability of the secondary structure of the RBD, and set a baseline for a comparison with the RBD structures affected by EFs.

We next performed EF-on simulations for different field intensities, namely EF= 10$^5$, 10$^6$, 10$^7$, 10$^9$ V m$^{-1}$, followed by the corresponding EF-off simulations. During the EF-on simulations, the secondary structure of the RBD was disrupted at multiple segments. Particularly, L3 undergoes a transition from the close structure with the two beta-sheets to an open and completely unstructured coil, reminiscent of L3 in SARS-CoV-1 [5,50] (figure 3a, right structure for EF=$10^7$ V m$^{-1}$). We evaluated the time evolution of the secondary structure of L3 (see Figure 3b for E=$10^6$ V m$^{-1}$ as example), which shows that beta-sheets gradually shrink during the first 200 ns of the simulation, until they finally deconstruct as turns or random coils before 1 $\mu$s. This indicates that the stretching of the protein by an external electric field leads to a destabilisation of the initial conformation. In the subsequent EF-off simulation, beta-sheets do not recover and L3 remains stable in its open unstructured state (Figure 3b). Coil or loop structures in proteins were previously described to have higher flexibility than highly ordered secondary structures such as $\beta$-sheets and helices[51]. We quantified the changes in flexibility of RBD by computing the root-mean-square-fluctuations (RMSF) of the RBD (Figure 3c). RMSF plots reveal that the EF modifies the flexibility of RBD inhomogeneously, with particular emphasis in the L3 loop and the RBM, in general. These results provide evidence that application of EF changes the secondary structure enduringly in segments that are critical for the interaction of the RBD with ACE2, and disrupts the spatial atomic organisation of the backbone and side-chain in key residues.

We also focused on the effect of EF in the spatial distribution of key residues of the RBM. We particularly analysed the residues that were previously described as participating in stabilisation of the RBM and in the interaction of the RBD with ACE2, since they contribute to the formation of a network of hydrogen bonds, hydrophobic and electrostatic interactions [52-53]. For instance, pairwise interactions between residues Cys488-Gly485 and Gln474-Gly476 have been pointed out as responsible for the stabilisation of L3[5]. In the no-EF simulation, the corresponding distances between residues were observed to remain within values around 4-5 Å that enable those interactions[54] (Extended Data, Fig 3). During EF-on, the same distances increase up to values between 8 Å and 10 Å, which causes a weakening of the inter-residue interaction. Another important set of residues of the RBM involved in hydrophobic contacts with the central region of the N-terminal helix of ACE2 are localised at the L2 and L3 loops, mainly comprising the aligned amino acids Leu 455, Phe456, Tyr473 and Tyr 489. In the original crystal structure 6M0J and in no-EF simulations, the aromatic residue from Phe456 is in close contact (less than 6 Å) with Tyr473 and the amino group of Lys 417, which leads to a very important stabilising internal $\pi$-cation interaction[53]. Application of an EF causes the following structural reorganisation of this set of residues: the increased mobility of L3 leads to a break of the Phe456 close-contact interactions, which reorders and misaligns this sequence of hydrophobic contacts (Extended Data, Fig.4, Fig.5). To complete the analysis of the EF-induced damage on the RBD, it is important to consider the residue Phe486, which plays a major role in the interaction of S with ACE2 because of its penetration into a deep hydrophobic pocket of ACE2 [43]. This interaction consists of a $\pi$-stacking of Phe486 with Tyr83 and two intermolecular contacts with the side chains of Leu79 and Met82 of ACE2, which contribute to L3 stability and the enhanced receptor binding. Figure 3d shows the RBM from the PDB structure 6M0J, from a representative conformation at EF=0 and after the EF-off run for EF=$10^6$ V m$^{-1}$. Notice that Phe486 is exposed in the β-turn of L3 in the absence of field, while the structure reorganisation by EF hides this residue in a persistent β-coil making it sterically inaccessible for the hydrophobic patch of ACE2. The above presented findings suggest that an EF induces disorder at sub-nanometre level that leads to unfavourable positions and orientations of important residues involved in the stabilisation of the RBM and the interaction with ACE2.

To further confirm that the atomic reorganisation in the S protein caused by the EF is likely to weaken its interaction with ACE2, we computed the electrostatic potential ϕ of the RBM for the crystalline structure and for the final structures of the no-EF and the EF-on runs, respectively. Electrostatic interactions have been intensively studied due to their importance in biomolecules recognition and binding [55,56]. We computed ϕ by solving the Poisson-Boltzmann equations for continuum electrostatics using the APBS package [57,58]. Figure 3e reveals, exemplifying for EF= $10^7$ V m$^{-1}$, that the spatial distribution of ϕ on the RBM is severely distorted upon EF application. In particular, the surface charge distribution in the L3 region is strongly affected. Notice that in the crystal structure (PDB ID 6MJ0) the binding surface on the ACE2 side exhibits a positive patch in the central region (blue area, contributed by residue Lys 31) that matches with the corresponding negative area on the RBD, bounded by a set of polar and acid residues including Glu471, Thr478, Glu484 and Gln493) [45]. These matching areas contribute to the strong electrostatic complementarity at the binding interface. After thermalisation (no-EF run) no significant changes occur in this region. However, upon rearrangement of L3, and especially the charged residues Glu471 and Glu484, due to EF application (EF-off run), the negative region in the RBD shifts and faces the negative part of ACE2, generating a repulsive force. At the same time, the region of the RBD opposite to Lys31 of ACE2 exposes non-polar residues. The same change in charge distributions is observed for other EF intensities (data not shown). These calculated electrostatic properties indicate that surface charge distribution in the RBM is strongly modified by EF and therefore the electrostatic complementarity between the S protein and ACE2, and therefore bonding between the RBD and ACE2 is disrupted (see Figure 3e as example for EF= $10^7$ V m$^{-1}$). Docking tests (see below) verified that these contacts are lost.

Finally, and in order to estimate the impact of positional and orientation changes of the residues along with charge and dipole rearrangements on the binding of RBD with ACE2, we performed calculations of the docking between the EF induced structures of the RBD and ACE2 using the tool PyDOCK [59]. We first listed the native-like contacts between RBD and ACE2 [5,45,52] in the 6M0J crystal structure (using a cutoff of 7 Å, see Methods and table in Extended data Fig.6). Then, we computed the remaining contacts from the above list in the best docked structures after EF and in absence of EF (including the final no-EF- and the 6M0J structures). We selected the cases with the highest number of preserved contacts (see Methods and Extended data Fig.6), which are shown in Figure 1c. The number of preserved contacts in the selected structures are taken as an estimate of how well the resulting structure of RBD can dock to the residues of ACE2, and therefore as a guess of how likely the RBD of S can bind to ACE2. The number of preserved contacts significantly decreases for increasing EF intensity (Figure 1c) up to the limit where all native contacts disappear for extremely high intensities. These estimates of docking efficiency further support our findings that EFs induce conformational changes disturbing the interaction of the RBD with ACE2 by spatially reorganising amino-acid dipoles and charges.

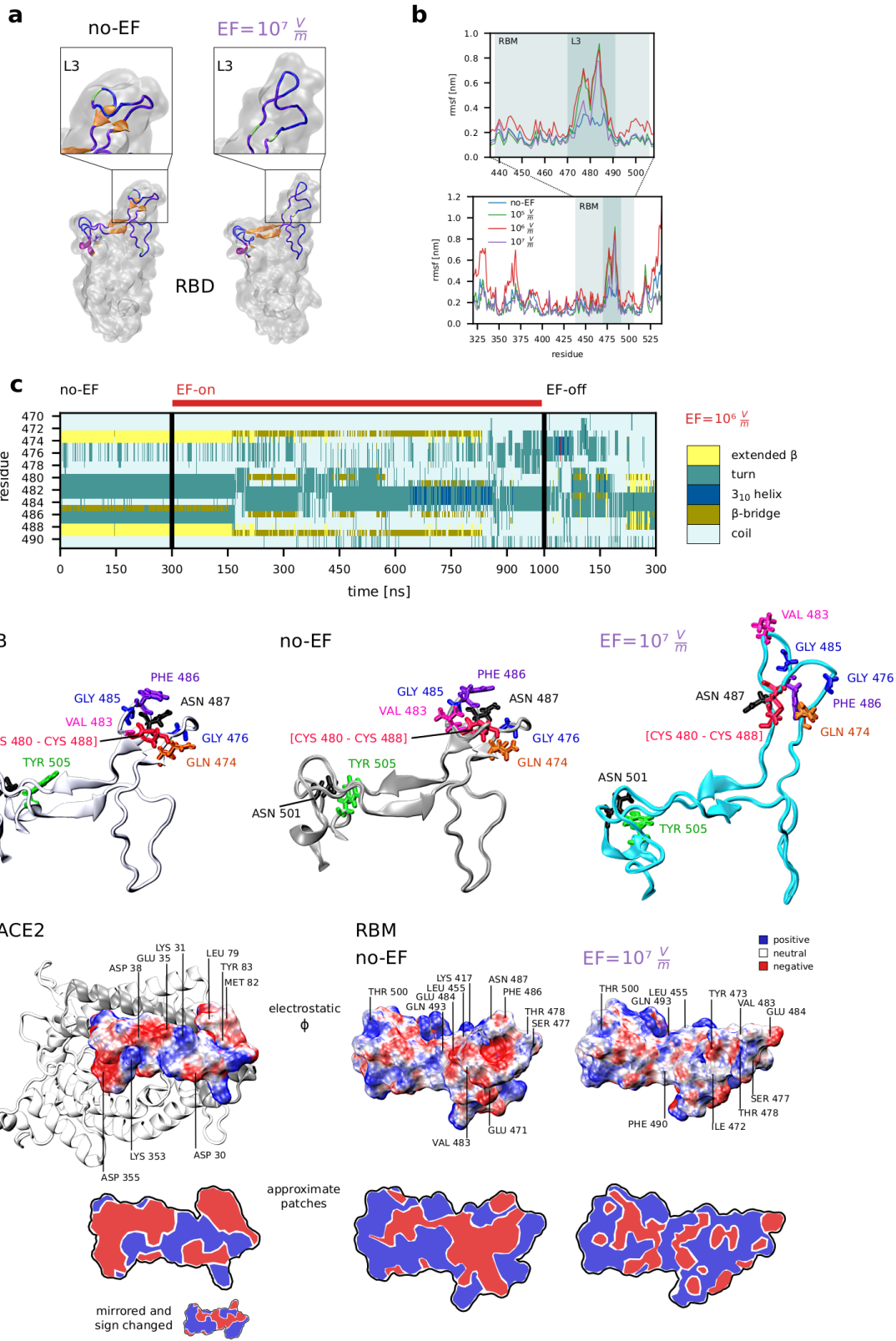

**Figure 3**: The secondary structure of the RBM can also be irreversibly perturbed by electric fields, affecting residues that participate in the binding to ACE2. **a,** The recognition loop L3 (Tyr470 to Pro491), exhibits two parallel $\beta$ sheets, which are responsible for a higher affinity to ACE2 [5,43]. The electric field induces a change of the secondary structure of L3 to an unstructured loop (example for EF=$10^7$ V m$^{-1}$). The spatial arrangement of key recognition residues is completely altered, which most likely inhibits their recognition function. **b,** Root mean square fluctuations of the amino acids of the RBD for different EF strengths. Residues in the receptor binding motif are highlighted and zoomed-in, showing the increased flexibility after EF application. **c,** Temporal evolution of the secondary structure of the L3 loop of the RBM (residues 470-491) for the no-EF, EF-on and EF-off runs (EF=$10^6$ V m$^{-1}$). The secondary structure is lost under field application and does not recover after EF switch-off. **d,** Close-up view highlighting the key residues of RBD participating in the binding with ACE2 for the crystal structure 6M0J (left), for a representative snapshot of the thermalised structure at 30 degrees Celsius (centre) and for a representative snapshot of the EF-off run (example for EF=$10^7$ V m$^{-1}$, right). **e,** Electrostatic potential at the surfaces of the relevant docking regions of ACE2 (left) and of the Receptor Binding Motif (RBM) at EF=0 (centre) and EF=$10^7$ V m$^{-1}$ (right). Red, white, and blue potential surface colours indicate negative, close to neutral, and positive charges, respectively. The lower panel shows a simplified planar visualisation of the positive and negative patches on both sides of the RBD-ACE2 bonding interface. The small inset shows a mirrored and sign-changed version of the charged surface of ACE2 in order to facilitate the appreciation of the charge complementarity between RBD and ACE2 in the absence of fields and its disruption upon EF application.

## Discussion

The structure of the S protein of SARS-CoV-2 seems to exhibit a finely tuned combination of geometrical, physical and chemical properties which provide efficiency in infectivity and bypassing the host immune system [15], as it is well documented in multiple studies on the effect of simple mutations, ligands, antibodies and recombinant protein expression systems[3,7,23,38,42,60,61].

In this work we showed that the application of relatively low to moderate static EFs can persistently change both the secondary and tertiary structures of S by rearranging and reorienting residues, thus disordering originally ordered segments through breaking and rebuilding of *hydrogen bonds and salt-bridges*. This results in reshaped local interactions and relative displacements between domains. Disruption of the secondary structure of S, particularly at the RBM, occurs under notoriously weaker EFs than those needed to produce significant changes in other proteins [26,28,32]. This suggests a possible causal link between structural vulnerability and affinity to ACE2. The structural features of S allowing the virus to avoid the immune response are in turn those ones particularly unprotected to EFs.

It is important to stress that EF of most of the intensities described here are achievable in practice in diverse contexts. For instance, EF strengths between $10^6$ V m$^{-1}$ and $10^7$ V m$^{-1}$, being well below the dielectric-breakdown threshold of water [62], are commonly used in industrial food processing to inactivate pathogens [63]. The immediate availability of cheap ways to produce EFs along with the basic skills required to manipulate them in both lab and industrial environments indicate that the effects described in this paper could spark the development of multiple solutions to mitigate the COVID-19 pandemic. For example, exposure of infectious microdroplets (aerosols) [64] or samples to EFs for timescales greater than a microsecond can generate inactivated or attenuated virions, and thereby cancel a mode of transmission that has been pointed out as the dominant for COVID-19 [64].

Our findings might also open up ways for EF based *in-vivo* (in the case of low EF-strengths such as $10^4$ V m$^{-1}$) and *in-vitro* therapeutic approaches. Recent works on exposure of mice to EF near to $10^4$ V m$^{-1}$ for several weeks as a treatment for diabetes reported no adverse side effects [65]. In the previous Section we showed that EFs disturb both the shape and charge complementarity to ACE2 and cause the exposure/hiding of key RBM-residues, leading to a dramatic reduction or suppression of the docking possibilities of S to the host cell. Owing to the same principle, the EF-induced motion of residues could be exploited to either favour or prevent the interaction of S with other molecules. For instance, cryptic epitopes (sequences that are inaccessible in the prefusion state) could be exposed under EFs, allowing for the binding of antibodies or ligands, as an alternative method to mutation-based techniques available in recombinant protein technology [66]. For instance, ligands acting as blockers can attach to regions of S and lock RBD up-down transitions [66]. Application of EF under the presence of blockers could yield to both disruption of functional motifs and locking dysfunctional conformations. The complete and ultrafast protein denaturation of the studied segment of S obtained under extremely high strengths ($10^9$ V m$^{-1}$, see Extended Data, Fig1) after less than 1 nanosecond suggests that intense EF pulses can cause analogous effects as antibodies that inactivate SARS-CoV-2 by premature conversion of S from pre-fusion to post fusion [21]. Although denaturation of S can also be achieved by raising temperature or changing the pH, the process is not controllable, in contrast to the nonthermal barrier-crossing occurring under EFs [67].

It is important to point out that our study shows consistent results between simulations based on different PDB-structures of S (6M0J and 6VSB). The driving forces governing atomic motion and residue rearrangement depend on dipole-alignment and charge distributions. Therefore, while single mutations can change the structural aspects of S, the driving forces under EFs at the local scale will be of the same order of magnitude throughout most of the sequence. Following this rationale, it is most likely that the recently discovered mutated types Variants of Concern 202012/01 (lineage B.1.1.7), 501Y.V2 (B.1.351) and P.1 (B.1.1.28) [68] with mutations affecting the RBD through residues N501Y, K417N, K417T and E484K will also be damaged by EFs of moderate strengths. There is preliminary evidence that the combination of these amino acid substitutions increases the transmissibility and confers resistance to several monoclonal antibodies, as well as resistance to neutralisation when using plasma from convalescent patients[69]. Consequently, these variants may increase the risk of infection or recently developed vaccines may be less effective. It is interesting to note that some of these multiple S protein mutations involve electric charge changes (from neutral to negative in A570D, from neutral to positive in P681H, from negative to positive in E484K and D1118H). This suggests that EF might be even more effective damaging some of the mutants.

Last but not least, and on the same line, the application of EFs to SARS-CoV-2, which strongly reduces its infectivity through modification of S, might also offer the possibility to generate, though applications in air filters or masks, partial immunity or cross reactivity against wild-type SARS-CoV-2 and its different mutations. Since there are some evidences that immunological memory due to infection with seasonal human coronaviruses (hCoVs) may generate cross-protection to SARS-CoV-2 [70-74], the EF-modified long-lasting states of the S proteins of SARS-CoV-2 might still provoke a certain immune response to the wild-type virus.

Summarising, this study demonstrates that EFs of different biologically relevant strengths change S of SARS-CoV-2 both at nanometre and sub-nanometre scales. Results of Fig. 2 show that the ensuing states under EF application clearly represent distinct atomic rearrangements depending on field strength. This raises the question whether tailored EF could be designed in order to drive S towards desired target structural states. Pulse trains, like those used in the food industry, or shaped oscillatory EFs of variable central frequency, envelope, duration and polarisation, could be optimised to promote a selective structural response in a similar way as in concepts involving electromagnetic fields [75].

**Acknowledgements**: Calculations for this research were conducted on the Lichtenberg high performance computer of the TU Darmstadt, the High Performance Center North (HPC2N) at SNIC, and at local dedicated workstations of our group. M.E.G and C.R.A acknowledge support by the PhosMOrg (P/1082 232) research unit of the University of Kassel. P.R. acknowledges the support of the Joachim Herz Foundation through its Add-on Fellowship for Interdisciplinary Life Science. C.R.A thanks the anti-Covid Consortium (Argentina) for valuable discussions. C.R.A. and P.R. thank Bernd Bauerhenne for his help with computing resources.

**Author contributions**: M.E.G. devised the main conceptual idea and the project. M.E.G, P.O-M and C.R.A. conceived the research plan. P.O-M. and C.R.A. pre-processed the files for simulations. C.R.A. and P.R. performed the MD simulations. C.R.A. performed detailed identification of changes in structures. C.R.A. and P.R. wrote the software for analysis and analysed the data. All authors analysed and discussed the results and their presentation. All authors wrote and corrected the paper.

## METHODS

### Protein structures preparation

The initial conformation (including atomic coordinates) was obtained from two available PDB structures with IDs: 6VSB and 6M0J, respectively. We considered in our simulations part of the chain A (residues 319 to 686) from 6VSB and the chain E (residues 333 to 526) from 6M0J, corresponding to the S protein and RBD, respectively. The missing hydrogen atoms and residues were added by using the CHARMM software (v. 43a1) [76] and CHARMM-GUI [77] with the CHARMM36 force field parameters [78-80] for A, and the tool Modeller [81] for E. Residues were protonated to fulfil pH 7 conditions and histidine (His) residues were treated as protonated on ND1 state (HSD). N-acetyl-β-glucosaminide (NAG) glycans were kept as in the original crystal structures and were modelled with the parameters of 2-acetyl-2-deoxy-$\beta$-D-glucosamine, with CONH fully charged atoms, using the same CHARMM36 force field.

### Molecular dynamics simulations

The simulations were performed using the GROMACS package (version 2019.4) [82-84]. CHARMM-36 force field parameters were adopted [78]. Both systems, 6VSB and 6M0J, were solvated with 298746 and 154770 TIP3P water molecules [85], respectively, with periodic boundary conditions. $Na^+$ and $Cl^-$ ions were further added to the boxes to simulate a salt concentration of 150 mM. The total number of atoms of segments A and E were 304841 and 158681, respectively. The LINCS algorithm was used to constrain all hydrogen bonds [86,87]. A cutoff of 12 Å was used for both the van der Waals and electrostatic interactions. The latter were computed with the help of the PME method [88,89] using a fourth order of cubic interpolation scheme with a grid size of 1.2 Å.

First, the energy of the system was minimised, while keeping heavy atoms at initial positions with harmonic constraints on backbone- and side chain atoms of 95.6 and 9.56 kcal/mol nm. Further restraints were applied on dihedral angles on atoms 3420, 3422, 3424, 3425, 3435, and 3439 with a force constant of 0.22 kJ/mol/rad$^2$ until the forces were less than 239 kcal/mol nm. This step was followed by a short equilibration stage of 1 ns using the NVT ensemble with the Nose-Hoover thermostat and a time constant coupling of 1ps. Then, and a longer equilibration run of 100 ns using the NPT ensemble was performed, in which the barostat was simulated using the isotropic Parrinello-Rahman algorithm[90] with a time constant coupling of 5 ps, a compressibility of 4.51x10$^{-5}$ bar$^{-1}$ and a reference pressure of 1 bar. The parameters of the thermostat used in both ensembles are the same.

The production runs were performed in the NPT ensemble. Once an equilibrated trajectory with no-EF was obtained, we used the atomic coordinates at 100 ns (PDB 6VSB) and 300 ns (PDB 6M0J) as the reference structures.

## Analysis of simulations

Unless otherwise stated, trajectory files were read and manipulated using GROMACS tools or the MDAnalysis Python library [91]. The VMD software [92] was used to visualise the MD trajectories and to draw the molecular representations.

<u>Principal Components Analysis (PCA) of the EF-on and EF-off trajectories</u>: In order to represent and visualise trajectories and states, we projected each of them in a suitable 2-dimensional space obtained by dimensionality reduction. Firstly, we computed and used dihedral angles as the generalised coordinates defining structural states, instead of the atomic cartesian coordinates. Dihedral angles, due to the local nature of its definition, are suitable to naturally separate internal motion from the overall motion of the protein. Furthermore, we transformed dihedral angles by splitting each one into two metric coordinates corresponding to its sine and cosine components [93]. This transformation from dihedral space to a linear metric space with a well-defined Euclidean distance, has been shown to preserve a unique representation while avoiding artifacts arising from the periodicity of angles [94]. Next, we found the reduced space by performing PCA over the above-mentioned metric coordinates and selected the first two components (corresponding to the highest eigenvalues) as the representative directions defining the reduced 2-dimensional space. PCA finds the directions of correlated motion by diagonalising the covariance matrix, with eigenvectors representing the directions of collective motion and eigenvalues ranked in descending order representing their amplitudes. We performed PCA using the scikit-learn Python library [95]. Then, we projected each trajectory onto the reduced PCA-space. Although we used the components corresponding the trajectory under EF = $10^7$ V m$^{-1}$ to project all results, similar plots arise if principal components corresponding to runs of other EF intensity are used. Therefore, the conclusions are independent of this choice.

Different residues and atomic distances (e.g. for disulfide bond and key amino acids) were calculated using customised Python scripts and the MDAnalysis library. Centroids of subdomains (for angle calculations) were computed as the arithmetic mean of the positions of the set of α-carbons of the corresponding residues. Using α-carbons speeds up calculations and yields negligible differences in the position of centroids with respect to the case where all atoms are considered. For distance between individual residues to analyse interaction, we computed the mean distance between all the atoms of each residue.

The changes on the secondary structure of the RBD over time with no-EF, EF-on and EF-off were estimated by the STRIDE algorithm implemented in the VMD software package version 1.9.4a38 (2019). The stride algorithm relies on hydrogen bond energy together with statistically derived backbone torsion angle data for the secondary structure characterisation in trajectories previously obtained by GROMACS.

<u>Free energy profile estimate</u>: the Landau free energy was estimated along the RMSD as a reaction coordinate using a path-sampling method[96,97,98] to approximate the potential of mean force (PMF) for each condition (no-EF, EF-on and EF-off) and for each EF strength. The free energy profile is then estimated by

$$F(rmsd) = -k_B T ln(\langle \delta(rmsd_k - rmsd) \rangle)$$

where $rmsd_k$ is the windowed RMSD value of the *k* position along the path, $k_B$ is the Boltzmann constant, *T* is the temperature and $\delta(...)$ is the Dirac delta function. Each path was binned using 20 windows along the RMSD coordinate.

**Electrostatic potential surface calculations**

The Adaptive Poisson-Boltzmann Solver (APBS) algorithm was used to calculate all potential maps[57] on the PDB 6M0J structural data and on selected frames from the MD trajectories. PDB formats were first prepared by PDB2PQR web server converted to PQR format using CHARMM force field with PROPKA set at pH = 7.0 [58]. Thereafter, we carried out the APBS analysis via Linearized Poisson-Boltzmann Equation in VMD software with settings parameters: solvent dielectric constant of 78.5, solvent radius of 1.4 Å, solute dielectric constant of 2.0, system temperature of 300 K, surface density 10.0 points/Å, and using harmonic average smoothing as surface definition.

**Molecular docking calculations**

Protein-receptor interactions were performed using pyDock[59] web server, which uses electrostatics and desolvation energy to score docking poses generated with FFT-based algorithms. We approached docking by first selecting the positions and orientations that optimise shape complementarity, followed by a rescoring based on electrostatic, van der Waals and desolvation energies [59]. The first 100 top-scoring structure complexes with the lowest total energy conformations were analysed to evaluate the binding interactions. This docking procedure was applied to the final structure after EF-off in each of the EF intensities evaluated in the shorter sequence. To quantify the likelihood of the docking, we first computed the contacts (as pairs of residues) between the RBD and ACE2 extracted from the 6M0J crystal structure of the RBD-ACE2 complex using a cutoff of 7 Å, measured as arithmetic mean of the atomic distances. We called these contacts native contacts, as they were described elsewhere to be essential for the binding between RBD and ACE2 [5,45,52]. From the resulting 100 best docked complexes for each input described above, we extracted the contacts following the same rule as for the native contacts, and selected the six structures that preserve the highest number of the native contacts. We accounted only for the first six structures to yield a comparable distribution of achieved preserved contacts between the different cases, since most of the 100 docked structures in each case showed zero preserved contacts and, therefore, would bias the distribution towards zero if they were included, leaving still recognisable tales but difficult to compare. Since we are interested in the likelihood of a correct binding to exist between a structure under scrutiny and ACE2, discarding the bulk of unmatched cases after the best ones is unlikely to lead to loss of important information about each of the specific structures under comparison. To validate the approach, we repeated the procedure consisting in docking, contacts computation and selection of the best contact preserving structures, for the RBD structures from 6M0J (separated from ACE2 and docked again) and the resulting after thermalisation. As shown above in the main text and Fig. 1, the computational docking of the structures taken from the 6M0J structure almost recover the exact position that was obtained experimentally in two of the cases. The case after thermalisation leads to less contacts than the case of experimental structure, but still around half of the contacts are found more than once, indicating an overall alignment between pairs of residues that are known to form bonds, and making a clear difference with respect to the analysed cases of structures that were exposed to EF.

## Reporting summary
Further information on research design is available in the Nature Research Reporting Summary linked to this paper.

## Data Availability
The raw simulation data that support the findings of this study are available from the corresponding upon reasonable request.

# Extended data

a)

b)

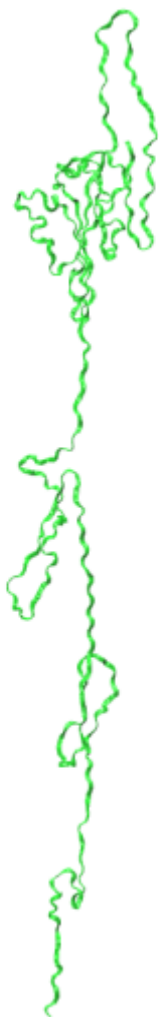
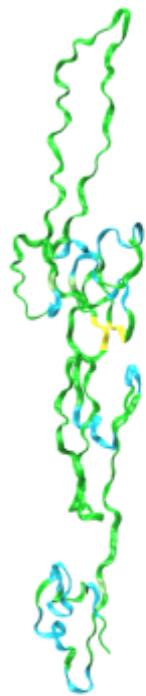

**Extended data. Fig. 1.** Snapshots of (a) the studied fragment of S PDB ID: 6VSB and (b) RBD PDB ID: 6M0J under an extreme and experimentally not very relevant EF of $10^9$ V m$^{-1}$, shown as a limiting case for the sake of comparison. Severe denaturation with complete loss of the secondary and tertiary structures of the protein within the first 1 ns takes place.

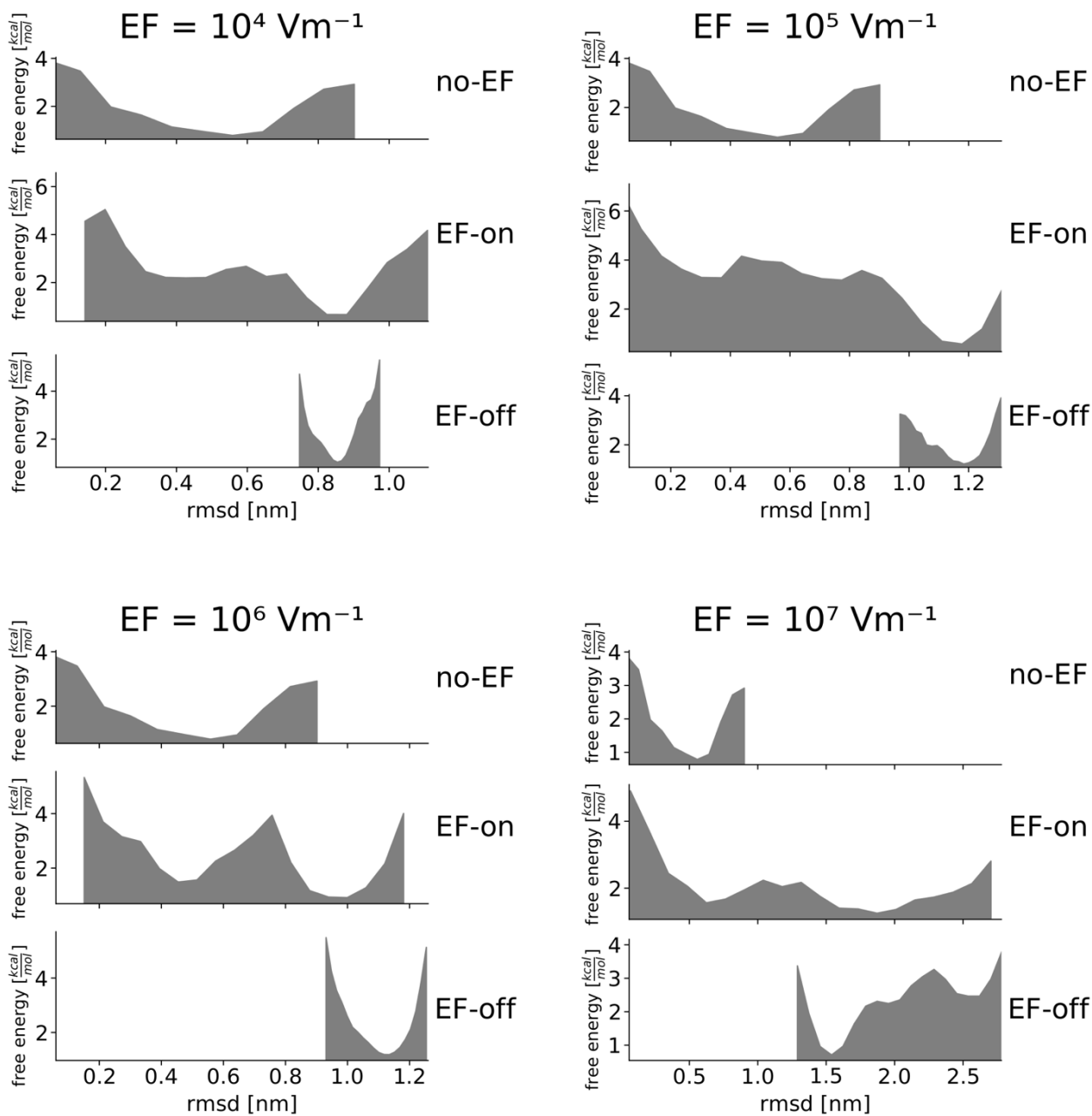

**Extended data. Fig. 2.** Estimated free energy profiles for no EF (top), EF-on (middle) and EF-off (bottom) corresponding to $10^4$ Vm$^{-1}$, $10^5$ Vm$^{-1}$, $10^6$ Vm$^{-1}$, $10^7$ Vm$^{-1}$. For all the EF strengths, the same qualitative landscape is observed, in which the protein state evolves until a new stable minimum is found and its basin is able to confine its further evolution. They also show that an energy barrier prevents a transition back to original conformation.

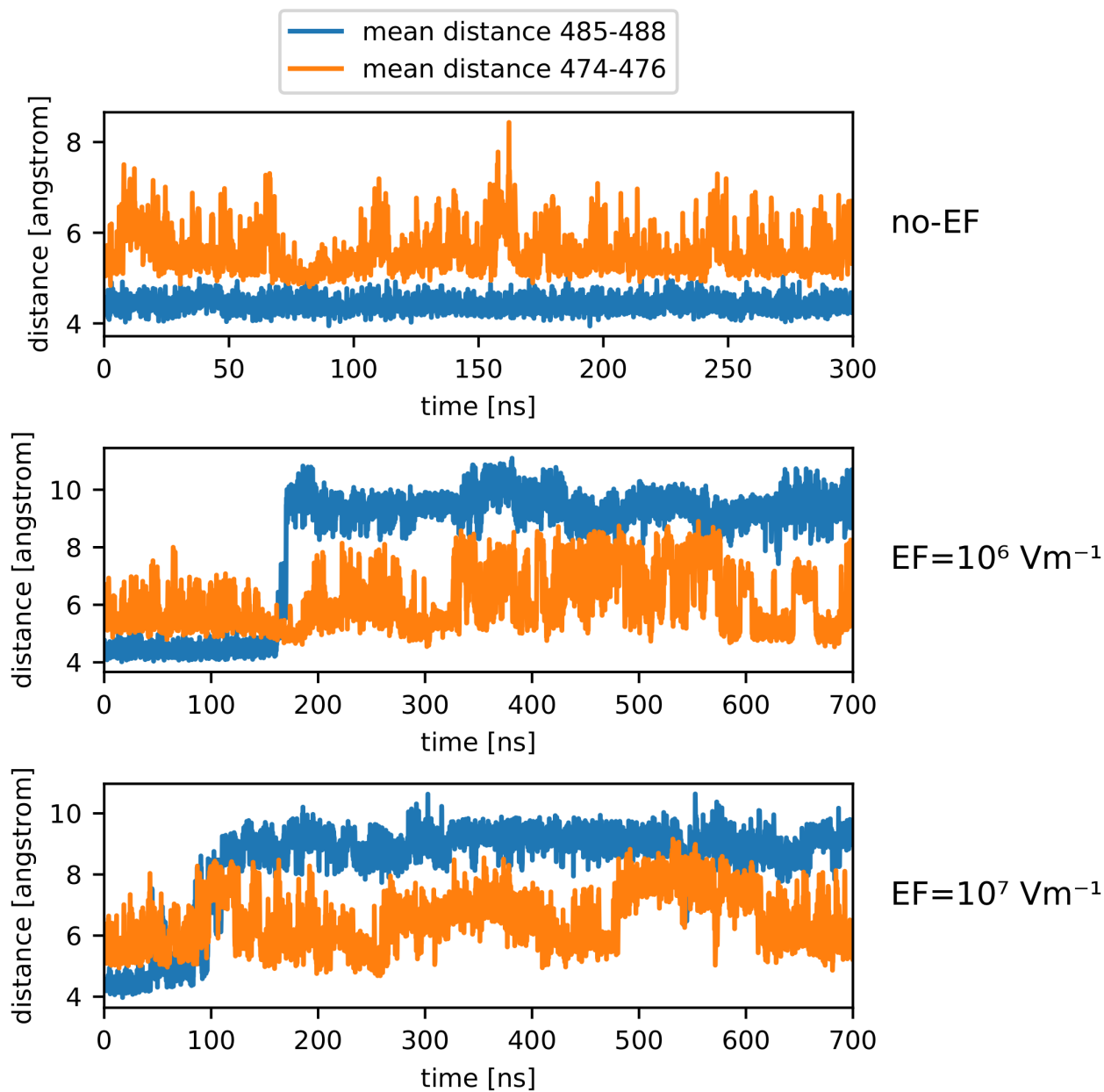

**Extended data. Fig. 3.** Time evolution of the mean distance between residues 485-488 and 474-476 under EFs for different strengths (no-EF and EF-on runs). Application of an external electric field has a significant effect on residues distance. In no-EF simulations, the corresponding distances between residues were observed to remain within values around 4-5 Å that enable these interactions. During EF-on, the same distances increase up to values between 8 Å and 10 Å, which causes a weakening of the inter-residue interaction.

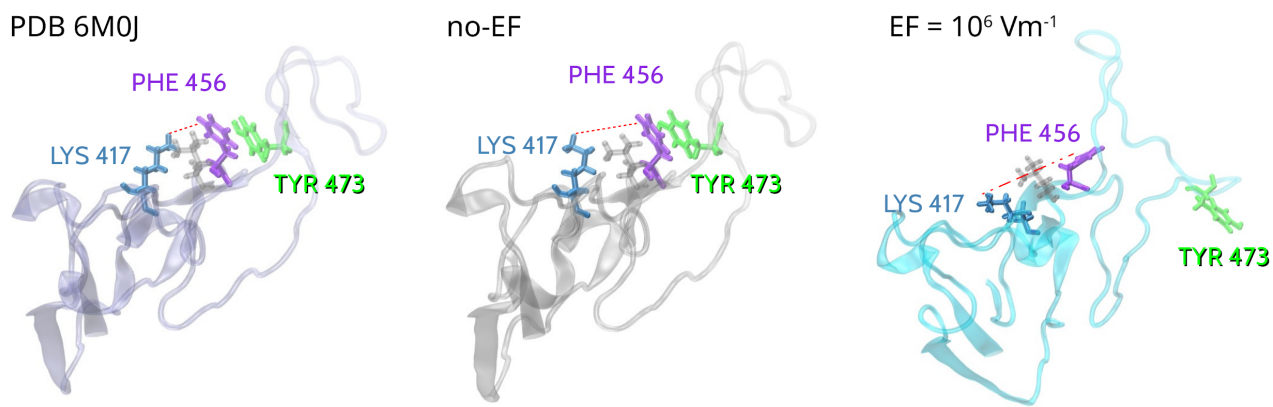

**Extended data. Fig. 4.** Snapshots of the RBM region for PDB ID 6M0J, EF=0 and EF=$10^6$ V m$^{-1}$ showing the contact interaction Phe456-Tyr473 and π-cation interaction between residue Phe456 and the side chain of Lys417. Application of an EF increased mobility of L3 leads to a break of these Phe456 close-contact interactions.

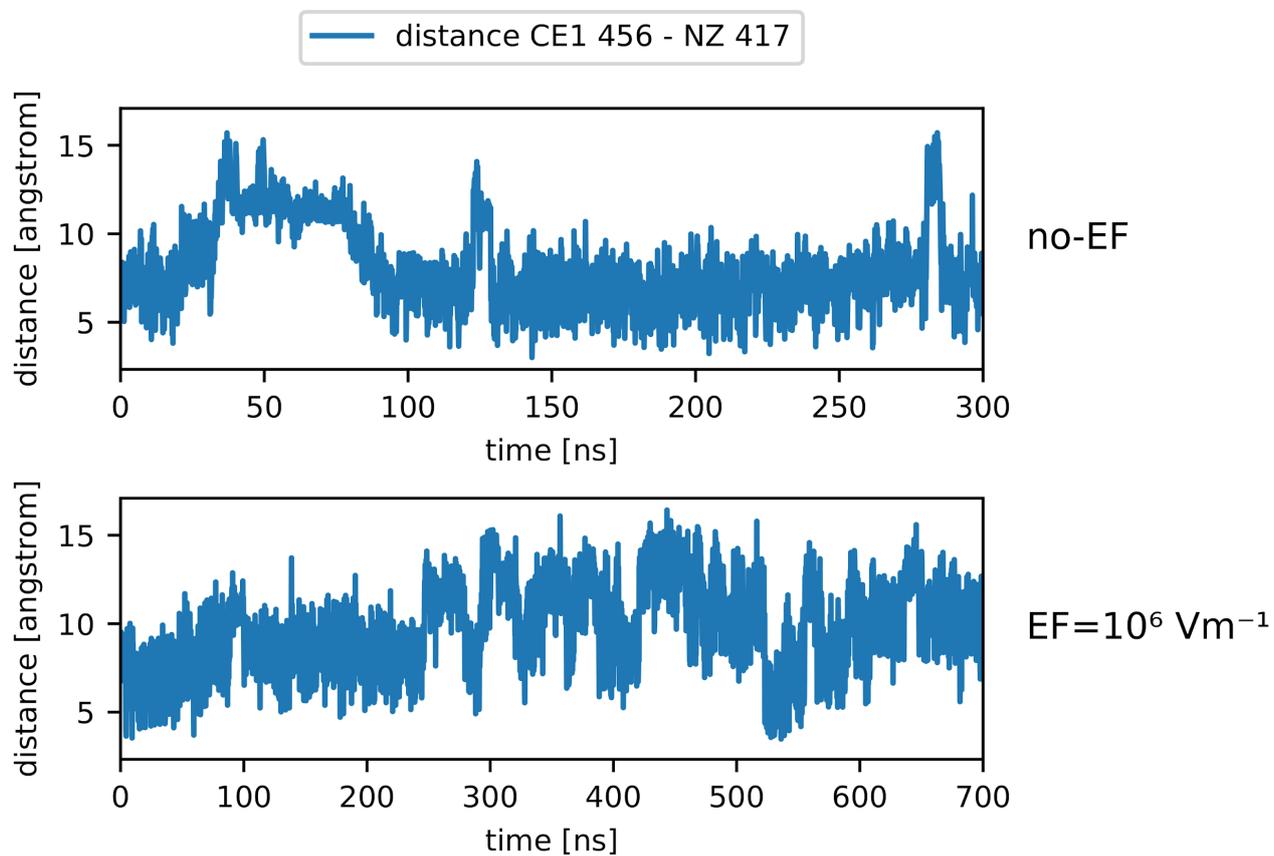

**Extended data. Fig. 5.** Time evolution of the mean distance between residues 417-456 under no-EF and EF-on ($10^6$ V m$^{-1}$) runs.

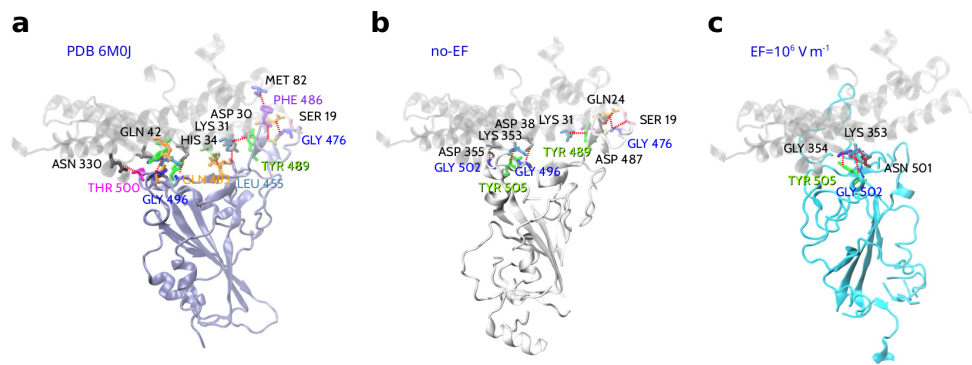

| Native contacts | | PDB ID 6M0J | | | | | | no-EF | | | | | | EF=10⁶ Vm⁻¹ | | | | | | EF=10⁷ Vm⁻¹ | | | | | |
|---|---|---|---|---|---|---|---|---|---|---|---|---|---|---|---|---|---|---|---|---|---|---|---|---|---|
| ACE2 | RBD | 1 | 2 | 3 | 4 | 5 | 6 | 1 | 2 | 3 | 4 | 5 | 6 | 1 | 2 | 3 | 4 | 5 | 6 | 1 | 2 | 3 | 4 | 5 | 6 |
| Ser19 | Gly476 | | | | | | | ■ | ■ | | ■ | | | | | | | | | | | | | | |
| Gln24 | Ala475 | ■ | ■ | ■ | | | | ■ | ■ | | | | | | | | | | | | | | | | |
| Gln24 | Gly476 | ■ | | ■ | | | | ■ | ■ | | ■ | | | | | | | | | | | | | | |
| Gln24 | Asn487 | ■ | | | ■ | | | ■ | ■ | | | | | | | | | | ■ | | | | | | |
| Tyr27 | Ala475 | ■ | ■ | | | | | ■ | | | | | | | | | | | | | | | | | |
| Asp30 | Leu455 | ■ | ■ | | | | ■ | | | | ■ | | | | | ■ | | ■ | | | | | | | |
| Lys31 | Tyr489 | ■ | | | | | | ■ | | ■ | | | | | | | | | | | | | | | |
| His34 | Gln493 | ■ | | | | | | | | | | | | | | | | | | | | | | | ■ |
| Asp38 | Gly496 | ■ | | | | | | ■ | | | | | | | | | | | | | | | | | |
| Tyr41 | Gln498 | ■ | | | | | | | | | | ■ | | ■ | | | | | | | | | | | |
| Gln42 | Gly446 | ■ | | ■ | | | ■ | | ■ | | | | | | | | | | | | | | | | |
| Met82 | Phe486 | ■ | ■ | | | | | | | | | | | | | | | | | | | | | | |
| Asn330 | Thr500 | ■ | ■ | | | | | | | | | | | | | | ■ | | | | | ■ | | ■ | |
| Lys353 | Asn501 | ■ | ■ | ■ | ■ | | | | | | | ■ | ■ | ■ | ■ | | | ■ | | | ■ | | | | |
| Lys353 | Gly502 | ■ | | | | | | | | ■ | | | ■ | ■ | | | | | | | ■ | | | | |
| Lys353 | Tyr505 | ■ | | | | | | | | ■ | | ■ | | | | | | | | | | | | | |
| Gly354 | Asn501 | ■ | ■ | ■ | ■ | | | | | ■ | | ■ | ■ | ■ | | | | | | ■ | ■ | ■ | | | |
| Gly354 | Gly502 | ■ | ■ | | | | | | | | | | | | | ■ | | | | | | ■ | | | |
| Gly354 | Tyr505 | ■ | ■ | ■ | | | | | | ■ | | ■ | | ■ | | | | | ■ | ■ | ■ | | | | |
| Asp355 | Thr500 | ■ | ■ | ■ | ■ | | | ■ | ■ | | | ■ | | | | | | | | | | | | | |
| Asp355 | Asn501 | ■ | ■ | | ■ | | | ■ | | ■ | | | | | | | | | | | | | | | |
| Asp355 | Gly502 | ■ | ■ | | ■ | | | ■ | ■ | | | | | | | | | | | | | | | | |

**Extended data. Fig. 6.** Snapshots showing RBD-ACE2 binding interface. **a.** Native contacts: the 22 contacts (as pairs of residues) between the RBD and ACE2 taken into consideration were extracted from the PDB ID 6M0J crystal structure of the RBD-ACE2 complex (cutoff of 7 Å). **b.,c.** Structures that preserve the highest number of the native contacts for EF=0 (9/22) and EF= $10^6$ V m⁻¹ (4/22). The orientation of the ACE2 binding interface is the same in all figures. **d.** Preserved contacts for the best 6 docking cases. Coloured cells represent contacts that are preserved. Native contacts involving residues from L3 in RBD (Tyr470-Pro491) are particularly affected when EF is applied.